\documentclass[usegraphicx]{mn2e}
\def\msun {{\rm M}_{\odot}}

\newcommand{\bm}[1]{\mbox{\boldmath$#1$}}

\begin{document}

  \title[QSO lensing magnification by galaxy groups]{QSO lensing magnification associated with galaxy groups}
  \author[A. C. C. Guimar\~aes, A. D. Myers and T. Shanks]
     {Antonio C. C. Guimar\~aes$^1${\thanks{E-mail: antonio.guimaraes@durham.ac.uk}},
       Adam D. Myers$^2$ and Tom Shanks$^1$\\
       $^1$ Department of Physics, University of Durham, Science Laboratories, South Road, Durham, DH1 3LE, U.K. \\
       $^2$  Department of Astronomy, University of Illinois, 1002 W Green Street,
       Urbana, IL 61801, USA }
     \date{Accepted 2005 June 23, Received 2005 June 21; in original form 2004 September 29}
     \maketitle
     
     \begin{abstract}
We simulated both the matter and light (galaxy) distributions in
a wedge of the universe and calculated the gravitational lensing
magnification caused by the mass along the line of sight of
galaxies and galaxy groups identified in sky surveys. 
A large volume redshift cone containing cold dark matter particles mimics
the expected cosmological matter distribution in a flat universe
with low matter density and a cosmological constant. 
We generate a mock galaxy catalogue from the matter distribution and identify
thousands of galaxy groups in the luminous sky projection. 
We calculate the expected magnification around galaxies and galaxy
groups and then the induced QSO-lens angular correlation due
to magnification bias. 
This correlation is an observable and can be used to estimate the average 
mass of the lens population and also make cosmological inferences. 
We also use analytic calculations and various analysis to compare the observational results with theoretical expectations for the cross-correlation between faint QSOs
from the 2dF Survey and nearby galaxies and groups from the APM and
SDSS EDR. 
The observed QSO-lens anti-correlations are stronger than the
predictions for the cosmological model used. 
This suggests that there could be unknown systematic errors in the 
observations and data reduction, or that the model used is not adequate. 
If the observed signal is assumed to be solely due to gravitational
lensing then the lensing is stronger than expected, due to more
massive galactic structures or more efficient lensing than
simulated. 
     \end{abstract}


\section{Introduction}

It is now very well established that mass concentrations on the
line of sight of distant sources act as gravitational lenses
distorting their image. One of the effects is the shear of
the image and that can be directly observed in drastic cases (strong
lensing), or statistically when the lensing is weak. The other
effect is the magnification of sources. Since lensing due to a
matter overdensity enlarges the solid angle of the source and
conserves its surface brightness, it can bring to view sources
that would be too faint to observe in a magnitude limited survey,
but at the same time it dilutes their population density. This
phenomenon receives the name of magnification bias, and the two
competing trends it engenders can give origin to both positive
correlation or anti-correlation between populations of objects
with very distinct redshift separation.

Several groups have measured background-foreground angular
correlation between populations of distant QSOs and nearby
galaxies or galaxy groups; see Bartelmann \& Schneider (2001) for
a review, and Guimar\~aes et al. (2001) for a compilation of
several results). 
Galaxy groups are particulary interesting because they trace 
higher density regions than galaxies in general, and therefore 
should yield a higher cross-correlation signal.
Croom \& Shanks (1999) found a lack of faint QSOs around 
galaxy groups and interpreted the
anti-correlation signal as due to gravitational lensing. Myers et
al. (2003) found that the lack of QSOs around galaxy groups
persists in larger samples, comparing QSOs taken
from the 2dF QSO Redshift Survey (Croom et al. 2004) to galaxies
taken from the Automated Plate Measurement (APM) galaxy survey (Maddox et al. 1990) and the Sloan Digital Sky Survey (SDSS)
Early Data Release (EDR; Stoughton et al. 2002). Using a simple analysis that
estimates an effective mass for the groups, both Croom \& Shanks
(1999) and Myers et al. (2003) find values that imply a high
density universe
($\Omega_m${\raise0.3ex\hbox{$\;>$\kern-0.75em\raise-1.1ex\hbox{$\sim\;$}}} 1 ).

Gazta\~naga (2003) finds a strong positive cross-correlation
between bright QSOs and galaxies from the SDSS EDR, to which he
suggests the interpretation of a large anti-bias ($b\approx 0.1$)
on small scales. A similar result is put forward by Myers et al.
(2005) from the strong anti-correlation found between faint 2dF
QSOs and APM and SDSS EDR galaxies.

Both interpretations -- high $\Omega_m$ or high anti-bias -- are
allowed by the underlying lensing theory, where the
background-foreground cross-correlation due to magnification bias
depends linearly on the mass density of the universe and on an
integration over the mass power spectrum. If we treat the two
possible explanations independently, the high $\Omega_m$
interpretation acts by increasing the overall lensing weighting
factor, whereas the low $b$ interpretation acts by increasing the
lens mass.
Of course both routes are not independent, since we
expect higher clustering for a denser universe. Nevertheless, the
results from these works are extreme, and at face value are in
discordance with other observations. Although the errors from the
referred works on the estimate of the universe matter density
parameter and galaxy bias are large, 
these results motivate a more elaborate analysis of the data.

An analytical theory of the background-foreground correlation due to 
weak gravitational lensing was developed by Bartelmann (1995), 
Dolag \& Bartelmann (1997), and  Sanz et al. (1997).
Guimar\~aes et al. (2001) incorporated into the formalism more 
elaborated galaxy biasing and used scale-dependent bias defined by 
the foreground population power spectrum.
M\'{e}nard et al. (2003) expanded the weak lensing approximation to second-order.
Jain et al. (2003) used the halo model to reformulate the bias dependence in terms 
of the halo occupation properties of the galaxy population.
Takada \& Hamana (2003) used the halo model and NFW profile to 
compute the full non-linear contribution to the cosmic magnification statistics.

The cross-correlation between galaxies or groups with QSOs offers a
direct opportunity to study and quantify gravitational
magnification and the masses of the foreground population of
lenses. 
We approached the problem by simulating
both the matter and light (galaxy) distributions, and 
directly calculating for each foreground lens traced by galaxies 
the angular magnification. 
We use a large-scale simulation of the large-scale structure of the
universe and a mock galaxy catalogue generated from the former to
compute the gravitational magnification due to the mass associated
with galaxies and galaxy groups. We use the observational results of Myers
et al. (2003 \& 2005) as study cases, that is, we seek to emulate the
parameters of these works in our simulations and calculations.

In Section \ref{matter-light} we describe how we generate a
simulated matter and galaxy distribution, how we identify groups
of galaxies in the mock catalogue, and test the simulation against
results from real galaxy surveys. In Section  \ref{lens_map} we
construct magnification maps for the simulated matter
distribution, and in Section \ref{Mag_centre} we calculate the
magnification around galaxy and galaxy groups, and the corresponding QSO-lens 
cross-correlation. In Section \ref{analytic} we present a simple 
analytic approach to the cross-correlation calculation and
compare it with the results from the mass-light simulation.

We show how the cross-correlation can be used to estimate the average
mass of the foreground lenses in Section \ref{mass_estim} and also compare
the mass estimated from the lensing results with the mass obtained
directly from the matter distribution simulation. We discuss our results 
in Section \ref{conclusion}.

\section{Mass and Light Distributions}
\label{matter-light}

\begin{figure}
  \centering
  \includegraphics[width=8.3 cm]{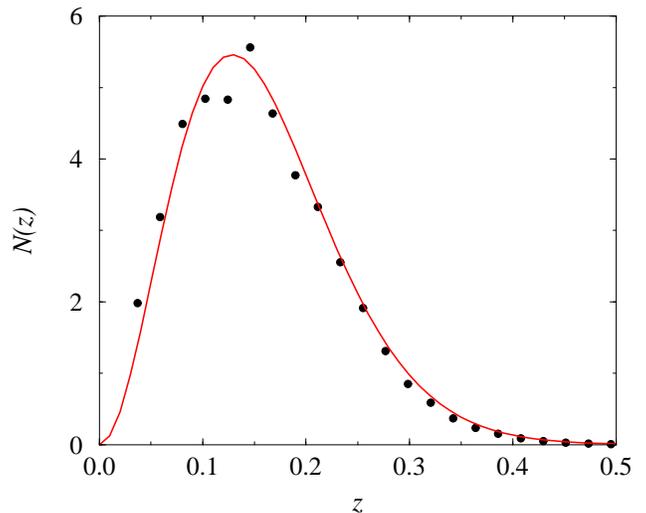}
  \vspace{-0.3cm}
  \caption{Galaxy redshift distribution. Circles are for the mock catalogue and the solid line is for expression (\ref{z_distrib}).  }
  \label{galreddist}
\end{figure}

We generated a simulation of the matter distribution in a 10 by 75 degree$^2$ segment of the universe centred at one observer at zero redshift and extending to $z=1$.
We use the output of the Hubble Volume Simulation (Frenk et al. 2000) that has $10^9$ particles of mass
$M_{part}=2.25 \cdot 10^{12}h^{-1}\msun$ in a periodic
$3000^3 h^{-3}\rm{Mpc}^3$ box.
We choose the ``concordance model'' simulation, which has
$\Omega_M=0.3$, $\Omega_\Lambda=0.7$, $\Gamma=0.21$, $\sigma_8=0.90$, initial fluctuations generated by CMBFAST, and force resolution of $0.1h^{-1}\rm{Mpc}$.

Our simulation of the matter distribution does not incorporate the evolution of the density fields, but this shortcoming is not very relevant for our purposes since we are interested in lensing by structures at small redshift ($z<0.3$).
Structures at larger redshifts, where evolution could be important, act mostly as noise for the lensing signal generated by the structures at small redshifts, since the large physical separation guarantees that they are uncorrelated.
One could incorporate evolution by using the light cone output of the Hubble Volume Simulation, but so far no one has created a galaxy mock catalogue from it.

We generated a mock galaxy catalogue using the simulated mass density field and adopting a bias prescription for the galaxy population.
We used the code of Cole et al. (1998), bias model 2, which is a 2-parameter model based on the final density field.
The mock catalogue generated has $4\cdot10^5$ galaxies magnitude limited to $B<20.4$, mean redshift $\bar{z}=0.15$, and redshift distribution displayed in
Figure \ref{galreddist}.
This distribution is well described by the expression for galaxy redshift distributions given by Baugh and Efstathiou (1993), with $\beta=1.5$ and $z_*=\bar{z}/1.412$:
\begin{equation}
  N(z) = \frac{\beta z^2}{z_*^3 \Gamma(\frac 3{\beta})}
  \exp{\left[-\left(\frac{z}{z_*}\right)^\beta \right]} \; .
  \label{z_distrib}
\end{equation}

Following Myers et al. (2003), we use the Turner and Gott (1976) algorithm
to select groups of galaxies from our mock galaxy catalogue.
The algorithm groups galaxies on the basis of their angular density on
the plane of the sky, ignoring the redshift coordinate.
About 45 per cent of groups with 7 or more
members identified by the Turner \& Gott algorithm have at least 7
members that are physically grouped (rather than being chance
alignments along the line of sight), though 95 per cent of ``Turner
\& Gott groups" with 7 or more members trace at least one or more
triplets of physically-grouped galaxies (Myers 2003).
Whether the groups identified by the Turner \& Gott algorithm are 
physically associated or not, they certainly always represent dense projections of galaxies along the line-of-sight, which should be hot-spots for lensing.

\begin{figure}
  \centering
  \includegraphics[width=7.55 cm]{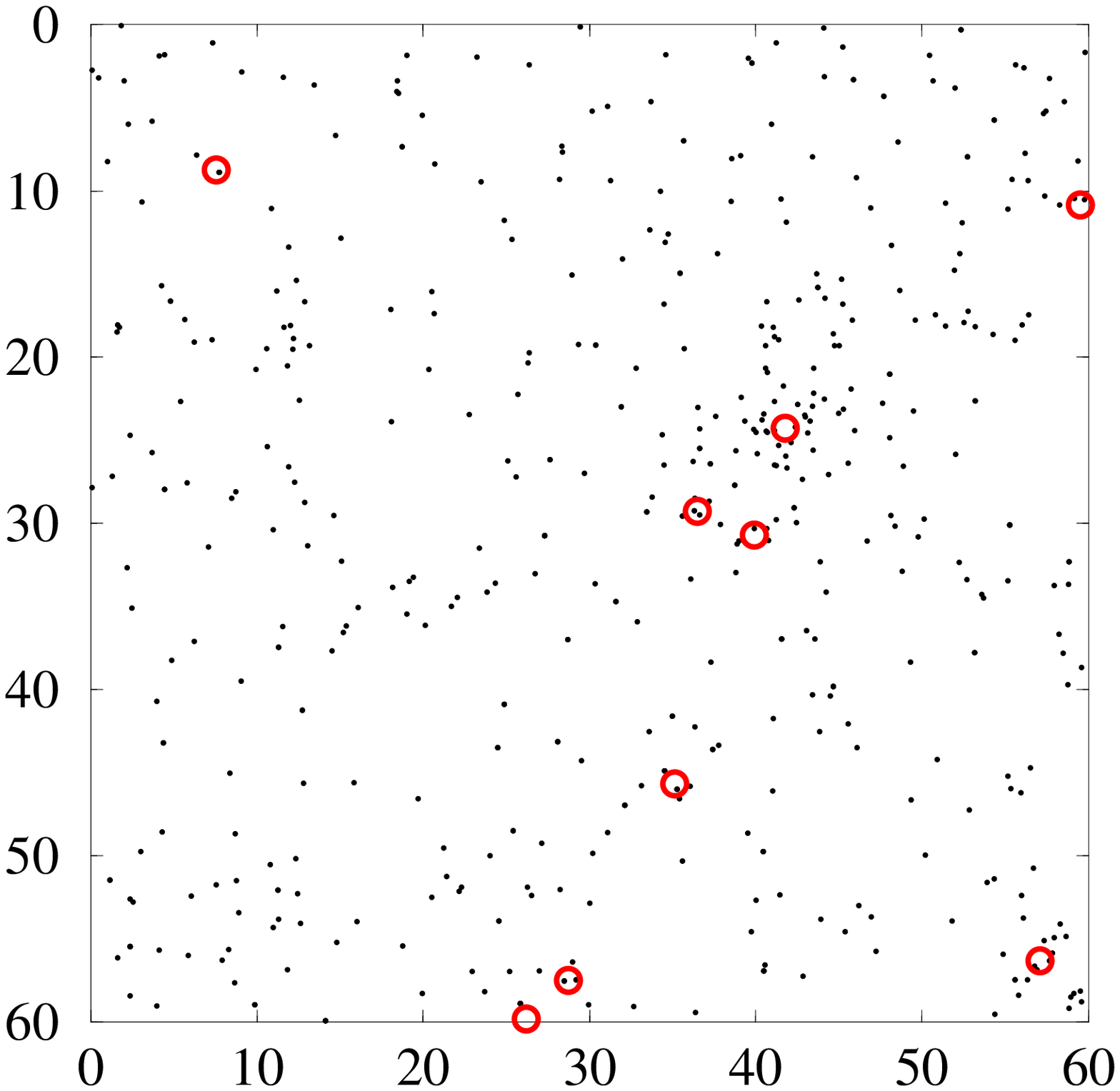}
  \includegraphics[width=7.5 cm]{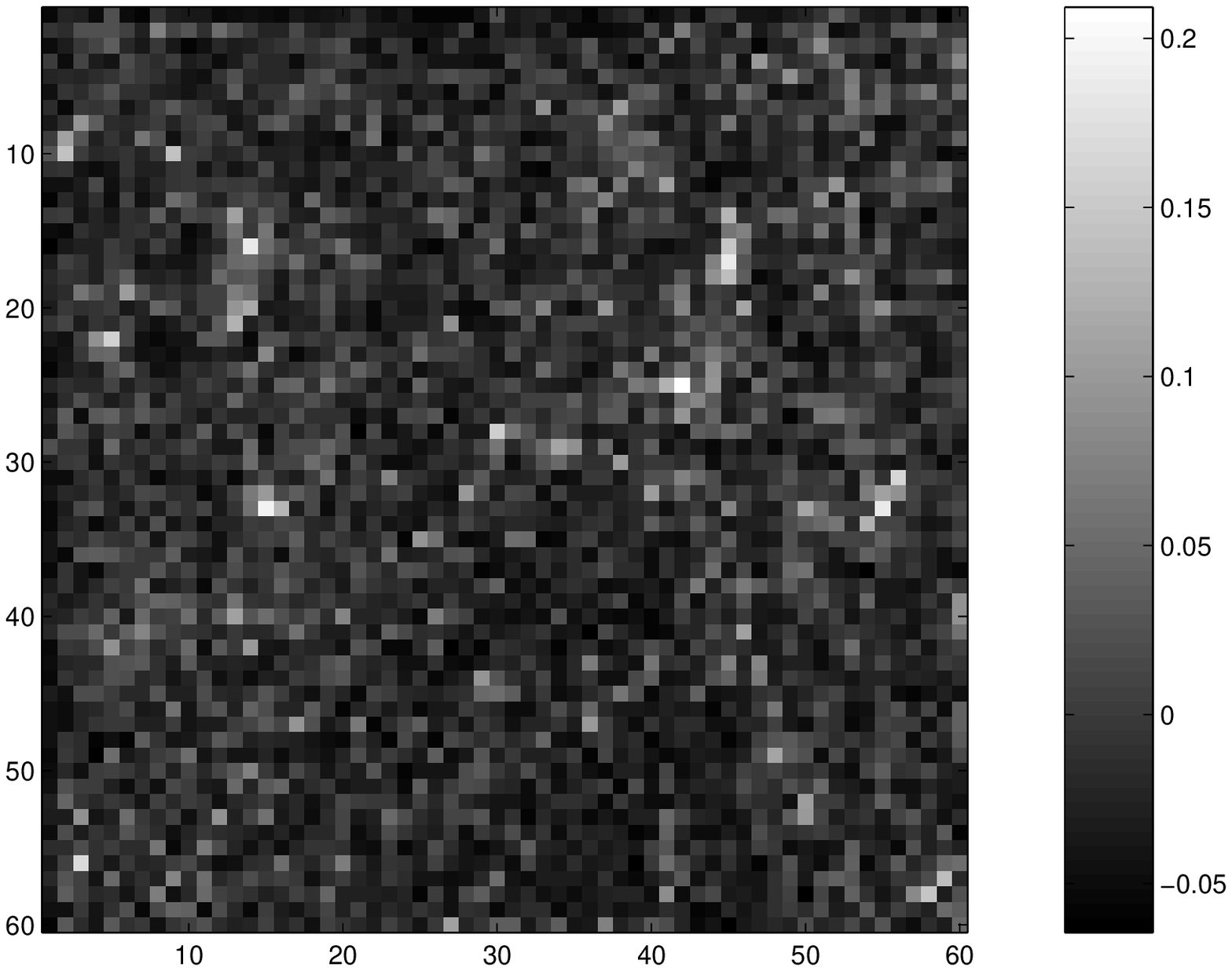}
  \vspace{0.2cm}
  \caption{Top panel: galaxies (dots) and groups with 7 or more members (circles indicate the group centre) in a 60$\times$60 arcmin$^2$ sky patch. Bottom panel: convergence map in the same region for a source population at $z=1$.}
  \label{mag_map}
\end{figure}

The top panel of Figure \ref{mag_map} shows a 1$\times$1 degree$^2$ projection of the simulated galaxy catalogue and the position of groups with 7 or more members identified in it.
One can observe some correlation between the position of groups and regions of high convergence (therefore magnification) in the lower panel of Figure \ref{mag_map}.
The next Section will discuss how the lensing map is generated.

Not all of galaxies that are present in this patch of sky shown in Figure \ref{mag_map} of the mock catalogue are visible in the projection because many occupy the same position.
This fact is a consequence of the low mass resolution of the N-body simulation.
Because the mass particles are very massive there are too few of them at small redshifts in relation to the number of galaxies.
Therefore the galaxy mock code attributes more than one galaxy to some mass particles, which explains why in the top panel of Figure \ref{mag_map} the groups of seven or more members and the overall sky patch appear to have fewer galaxies than expected.

\begin{figure}
  \centering
  \includegraphics[width=8.3 cm]{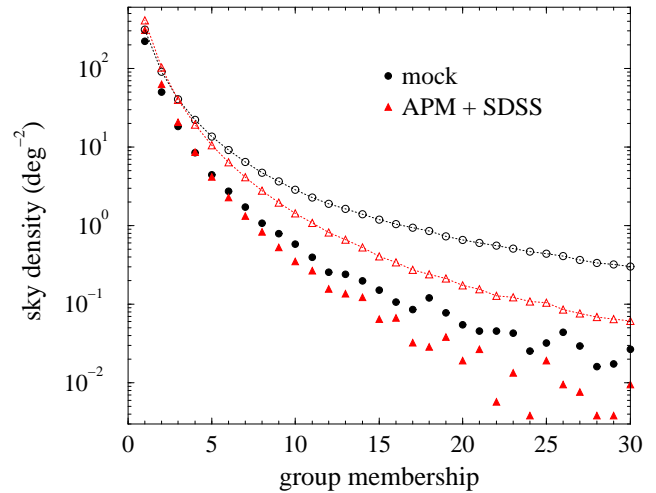}
  \vspace{0.cm}
  \caption{Sky density of galaxy groups identified in 2D for mock catalogue and APM + SDSS. Filled symbols are for simple membership (exactly $N$ galaxies), and open symbols connected by lines are for groups with {\it at least} $N$ galaxies.}
  \label{groupmemb}
\end{figure}

The low mass resolution, and the group identification algorithm based on friend-of-friends, which identifies neighboring galaxies inside a given radius, induce a higher identification of groups than would be expected.
Assigning the same coordinates to more than one galaxy may yield a false group identification.
Figure \ref{groupmemb} shows the group density on the sky as a function of group membership.
The sky density for groups with 7 or more members is $6.5$deg$^{-2}$, which is larger than that obtained from real data (APM + SDSS), $3.8$deg$^{-2}$.
This discrepancy may be relevant for the objectives of this paper and reveals that the galaxies in the mock catalogue form more groups than what is observed, even though they have comparable sky density (540 deg$^{-2}$ for observed galaxies and 530 deg$^{-2}$ for mock galaxies).
One way to compensate for this discrepancy is to assume that the real group population is the one with same sky density as the observed. 
Mock groups with 9 or more members have this property as compared to observed groups with 7 or more members, and we will use those in some of or calculations, however the difference is small.

\begin{figure}
  \centering
  \includegraphics[width=8.3 cm]{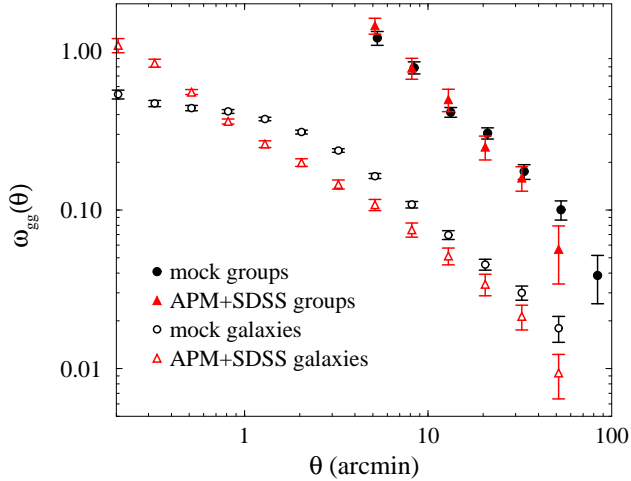}
  \vspace{0.cm}
  \caption{Angular auto-correlation for galaxies and galaxy groups with 7 or more members. Errors are field-to-field.}
  \label{groupcorrel}
\end{figure}

Figure \ref{groupcorrel} shows the angular auto-correlation for galaxies and groups with 7 or more members, comparing the mock catalogue with a combination of the APM and SDSS results.
Mock galaxies have a lower angular auto-correlation than real galaxies for separation less than 1 arcmin (below the simulation resolution), and a somewhat higher one for larger separation.
Groups with 7 or more members identified in the mock catalogue have comparable clustering to groups detected in the APM survey and SDSS.

\section{Lensing Maps}
\label{lens_map}

The full knowledge of the mass distribution between the observer at $z=0$ and a source distribution at high redshift allows us to calculate the gravitational lensing of sources by the intervening mass inhomogeneities.

In the cosmological context, and assuming the Born approximation, the convergence can be calculated by
\begin{equation}
  \kappa({\mbox{\boldmath$\theta$}}) =
  \int_0^{y_{\infty}} { W(y) \delta({\mbox{\boldmath$ \theta$}},y) dy} \; ,
  \label{convergence2}
\end{equation}
where $\delta$ is the density contrast, $y$ is a comoving distance, 
$y_{\infty}$ is the comoving distance to the horizon, and $W(y)$ is a lensing weighting function
\begin{equation}
  W(y) = \frac{3}{2} \left( \frac{H_o}{c} \right)^2 \Omega_m
  \int_y^{y_{\infty}} { \frac{G_q(y^\prime)}{a(y)}
  \frac{f_K(y^\prime-y)f_K(y)}{f_K(y^\prime)} dy^\prime } \; .
  \label{weight}
\end{equation}
$G_q$ is the source distribution, $a$ is the scale factor, and $f_K$ is the curvature-dependent radial distance.

Since our simulation of the mass distribution consists of a discrete set of mass points, equation (\ref{convergence2}) can be more suitably written for computational purposes as
\begin{equation}
  \kappa({\bm \theta})= 
  \sum_{i=1}^N{ \frac{\Sigma({\bm \theta}_i,z_i)}{\Sigma_{cr}(z_i)} } 
  - \kappa_{min}\;.
  \label{convergence1}
\end{equation}
The sum is perfomed over all mass particles in the simulation falling in the sky patch covered.
$\Sigma({\bm \theta}_i,z_i)$ is the surface density of one mass particle in one angular cell of the grid adopted and
\begin{equation}
  \Sigma_{cr}=\frac{c^2}{4\pi G} \frac{D_s}{D_d D_{ds}}   \;,
\end{equation}
where $D_s = a(y_s) f_K(y_s)$ is the angular-size distance to the source 
(our maps assume a source surface at $z=1$), 
$D_d= a(y_d) f_K(y_d)$ is the angular-size distance to the mass particle (deflector), and $D_{ds}= a(y_{s}) f_K(y_{s}-y_{d})$ is the angular-size distance from the source to the mass particle.
The empty beam ($\delta({\mbox{\boldmath$ \theta$}},y)=-1 $) value for the convergence in expression (\ref{convergence2}) defines $\kappa_{min}$, which can be also obtained from the ensemble average of the terms of equation (\ref{convergence1}) if one uses that $\left< \kappa ({\bm \theta}) \right> = 0$.

The method differs from the multi-plane method in that it does not divide the space in cubes and does not project the mass particles in planes. Therefore it does not suffer from discontinuity problems across space as is the case with the multi-plane approach, and also avoids the cumulative effective smoothing generated by the multi-plane grids.
The Born approximation is justified for the lensing systems and resolution level that we use here (see Jain, Seljak and White 2000, and Vale and White 2003).

The resolution of the convergence map is determined by its grid size 
$\theta_{grid}$, and the effective angular resolution of the projected mass simulation $\theta_{mass}$, which can be estimated using a similar prescrition used by M\'enard et al. (2003) to calculate the effective smoothing scale
\begin{equation}
\theta_{mass} = \int{W(y) \frac{l_{mass}}{D_{ang}(y)} dy} \;,
  \label{ang_res_LSS}
\end{equation}
where $W(y)$ is the normalized weighting function (\ref{weight}), $l_{mass}$ is a linear resolution of the mass simulation, and $D_{ang}(y)$ is a angular-size distance.
Assuming $l_{mass}$ to be given by the force resolution of the Hubble Volume Simulation, we obtain $\theta_{mass}=0.5$ arcmin, which is less then the grid size of the convergence map $\theta_{grid}=1$ arcmin. 

The convergence map produced has minimum value $\min[\kappa (\theta)]=-0.065$, and maximum value $\max[\kappa (\theta)]=0.92$, which indicates that we are probing regions where departures from the weak lensing regime, $\kappa \ll 1$, may be non-negligible.

The magnification map is related to the convergence $\kappa$ and shear $\gamma$ by
\begin{equation}
  \mu({\bm \theta}) = 
  \frac{1}{\left| \left[1-\kappa({\bm \theta}) \right]^2-\gamma^2({\bm \theta}) \right| } \;.
  \label{magnification}
\end{equation}
The convergence and shear fields have some common statistical properties, for example $\left< \gamma^2({\bm \theta}) \right> = \left< \kappa^2({\bm \theta}) \right>$, but this identity is not valid locally. 
Although ensemble averages of convergence and shear fields cannot be rigorously used inside expression (\ref{magnification}) for the calculation of statistical properties of the magnification, one can expand expression (\ref{magnification}) in powers of $\kappa$ and $\gamma$ and them use the ensemble averages of these fields to calculate approximately the ensemble average of $\mu$ and powers of it.
This is the approach used, for example, by M\'{e}nard et al. (2003) to calculate magnification correlations to second-order approximation.
Other approach is to use local approximations, and since we are mostly interested in the magnification around identified luminous lenses (galaxies and groups), we use this avenue. This will become clearer in the next Section.

One approximation (Fan and Chiueh 2000) is to assume the shear to be negligible in expression (\ref{magnification}),
\begin{equation}
  \mu = \frac{1}{(1-\kappa)^2 } \;,
  \label{magnification2}
\end{equation}
In fact, for the magnification generated by a NFW or SIS profile the shear will only become important in expression (\ref{magnification}) very close to the centre, or below than the 1 arcmin scales that we can probe in our lensing map given the relative poor resolution of our mass simulation on small scales.

For a SIS profile the shear around the centre of the mass distribution is equal to the convergence at the same point
\begin{equation}
\kappa_{SIS}(\theta) = \gamma_{SIS}(\theta) 
= \frac{2\pi\sigma^2_v D_{ds}}{c^2 D_s} \frac{1}{\theta} \;,
\end{equation}
which implies a magnification
\begin{equation}
  \mu = \frac{1}{\left| 1-2\kappa \right| } \;.
  \label{magnificationSIS}
\end{equation}
More generally, for any circularly symmetric profile one could use that the shear is given by
\begin{equation}
\gamma(\theta) = {\bar \kappa}(< \theta) - \kappa(\theta) \;,
\label{tangential_shear}
\end{equation}
where ${\bar \kappa}(< \theta)$ is the average value of the convergence inside a radius $\theta$.

For our simulated groups with 9 or more galaxies the modulus of the average shear in ring around the centre of the group is approximately $1/3$ of the average convergence in the same ring. 
Therefore for scales that we can probe the effective magnification (\ref{magnification}) falls between approximations (\ref{magnification2}) and (\ref{magnificationSIS}), being closer to the former.

The weak lensing approximation, $\kappa \ll 1$, is a first order expansion of expression (\ref{magnification}),
\begin{equation}
  \mu = 1+2\kappa  \;.
  \label{magnification3}
\end{equation}

We calculated the magnification using various approximations, but limit to report in our plots the results from expressions (\ref{magnification2}) and (\ref{magnification3}), which already allows us to see some departure from the weak lensing linear regime.
Expression (\ref{magnification3}) directly yields a null mean field for the magnification map since $\left< \kappa ({\bm \theta}) \right> = 0$, but the same does not occur with expression (\ref{magnification2}).
It is necessary to renormalize the magnification map so that
$\left< \mu ({\bm \theta}) \right> = 1$.
One can impose this condition by the means of two transformations
\begin{equation}
  \mu({\bm \theta}) \rightarrow \mu({\bm \theta}) / \bar{\mu} \;,
\end{equation}
or
\begin{equation}
  \mu({\bm \theta}) \rightarrow \mu({\bm \theta}) - (\bar{\mu}-1)  \;,
\end{equation}
where $\bar{\mu}$ is the mean value of the non-normalized magnification map.
For our simulation, and using approximation (\ref{magnification2}), we obtained $\bar{\mu}=1.0051$, and the difference between the two renormalizations is negligible for our results here.

The importance of considering departures from the weak lensing approximation has been highlighted by Barber and Taylor (2003), Takada and Hamana (2003), and M\'{e}nard et al. (2003).

\section{Magnification Around Group Centres and QSO-Group Cross-Correlation}
\label{Mag_centre}

\begin{figure}
  \centering
  \includegraphics[width=8.3 cm]{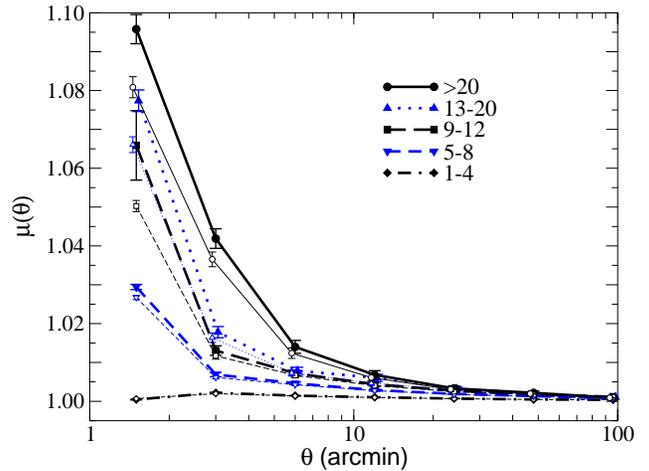}
  \vspace{0.1cm}
  \caption{Average magnification around mock galaxy groups. The number ranges in the legend are the number of galaxies for the sets of groups. Thick lines with filled symbols include departures from the weak lensing approximation, and thin lines with open symbols are the weak lensing approximation for the magnification calculation. Errors are the standard deviation of the mean.}
  \label{magXtheta}
\end{figure}

We used our magnification maps and the galaxy groups identified in the galaxy mock to calculate the average magnification in annular regions around group centres -- see Figure \ref{magXtheta}.
One can observe that the weak lensing approximation underestimate the average magnification, and that there is a strong dependence of the magnification on group membership.
Error bars are standard deviations from the mean value considering total sampling, i.e. all cells from the grid are probed.
A more realistic error estimate for an observational setting would consider a more sparse sampling.
A rough estimate is to assume that the effective sampling used on the simulations is given by the density of cells (1 ${\rm arcmin}^{-2}$ for our grid), and in a real observation it is given by the background source density.
This would yield for the study case adopted here, which has a QSO sky density
$\rho_{QSO}=43 {\rm deg}^{-2} $, a factor $\sqrt{\rho_{\rm cell}/\rho_{QSO}}=9$ larger for the magnification error bars.

The magnification alters the light flux $S$ from sources, which implies that if $\mu>1$ a source that could be too faint to be observed in a magnitude-limited survey may be brought to view.
The number density of sources in the magnitude limited survey is $ N(>S/\mu)$.
But at the same time the magnification also alters the area behind a lens,
$A^\prime=A/\mu$, so in a $\mu>1$ region the area behind the lens is expanded and the source counting density is diluted.
These two competing effects are compared in the net enhancement factor
\begin{equation}
  q\equiv \frac{A^\prime N(>S/\mu)}{A N(>S)}  \; .
\end{equation}
If the source cumulative number counts by flux is of the form $N(>S)\propto S^{-s}$ then the enhancement factor reduces to $q=\mu^{s-1}$.
The coefficient $s$ of the number-flux relation relates to the coefficient of the number-magnitude relation by $s=2.5\beta$. We use $\beta=0.29$ where required (Myers et al 2003).

The enhancement can be estimated as the ratio of the the observed number of QSO-group pairs, $DD(\theta)$, to the expected number of random pairs, $DR(\theta)$.
If we recall the cross-correlation estimator
\begin{equation}
  \omega_{qg}(\theta) =  \frac{DD(\theta)}{DR(\theta)} -1  \; ,
  \label{estimator}
\end{equation}
then
\begin{equation}
  \omega_{qg}(\theta) = \mu(\theta)^{s-1} - 1  \; .
  \label{w_qg-mag-rel}
\end{equation}

\begin{figure}
  \centering
  \includegraphics[width=8.3 cm]{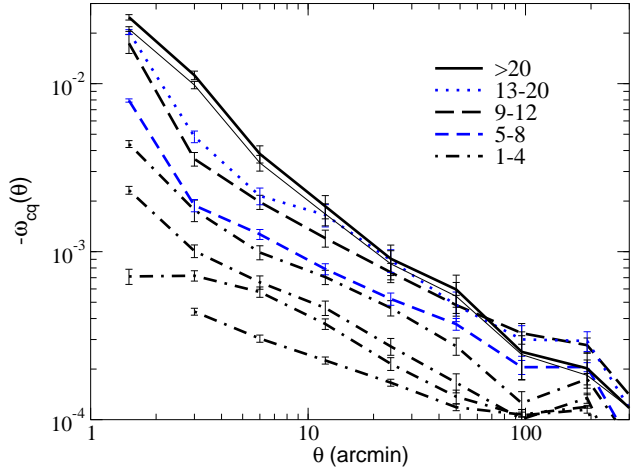}
  \vspace{0.2cm}
  \caption{Cross-correlation between QSO and mock galaxy groups 
(we plot $-\omega_{qg}$ to allow the use of logarithmic scale). 
The number ranges in the legend are the number of galaxies for the sets of groups. Thick lines include departures from the weak lensing approximation, and thin lines are the weak lensing approximation for the magnification calculation. Curves for groups of membership 1 to 4 are shown individually (from bottom to top for low to high membership). Errors are the standard deviation of the mean.}
  \label{cross-correl}
\end{figure}

Figure \ref{cross-correl} shows the QSO-group correlation function for groups with different galaxy membership.
A departure from weak gravitational lensing is illustrated for groups with more than 20 members.
The curves show a strong dependence in relation to group membership, indicating its relation to the underlying mass overdensity.

\begin{figure}
  \centering
  \includegraphics[width=8.3 cm]{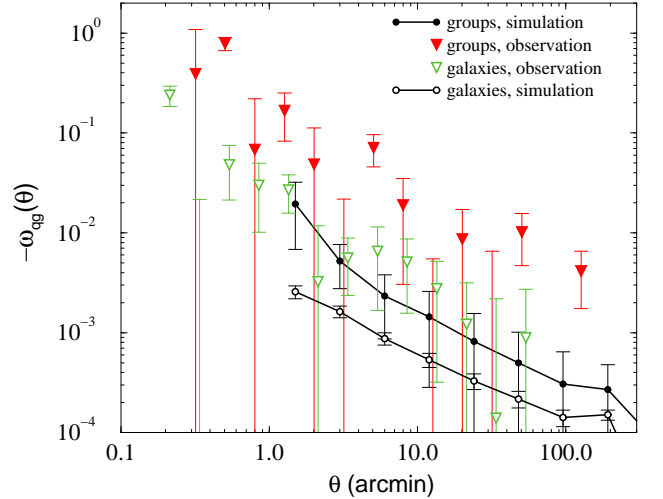}
  \caption{QSO-galaxy and QSO-group cross-correlation.
Observational data is from Myers et al. 2003 and 2005 with field-to-field errors.
Simulation uses groups with 9 or more galaxies and estimated error with same source density as observed data. Some observational points fall below the shown logarithmic scale}
  \label{red-cross-correl}
\end{figure}

Figure \ref{red-cross-correl} compares the cross-correlation results from simulations to data from Myers et al. (2003) for groups, and Myers et al. (2005) for galaxies.
Simulation results for angles smaller than 1 arcmin cannot be obtained due to limited simulation resolution; however for angles from 1 to 100 arcmin the comparison with data gives a large disagreement between the amplitudes of observed and simulated cross-correlations.
We find the parameters of a power law that best describe the observed and simulated QSO-galaxy and QSO-group cross-correlations, and quantify the disagreement in amplitude between the observed and simulated results, which is larger for groups (factor of $\sim$20) than for galaxies (factor of $\sim$7).

\section{Simple Analytic Approach}
\label{analytic}

In this section we present a simple analytic approach to check and help the interpretation of the results from simulation.

Expression (\ref{estimator}) for the cross-correlation is equivalent to (see Appendix \ref{cross-corr-def-est})
\begin{equation}
  \omega_{qg}(\theta) \equiv
  \left< \left[ \frac {n_q(\bm{\phi})}{\bar{n}_q} -1 \right]
  \left[ \frac {n_g( \bm{\phi}+ \bm{\theta})}{\bar{n}_g} -1 \right]
  \right>  \; ,
  \label{cross-correl-def}
\end{equation}
where $n_q$ and $n_g$ are the QSO and galaxy (or galaxy group) densities (a bar over a quantity indicates its mean value), and $\left< ... \right>$ represents the average over $\bm{\phi}$ and the direction of  $\bm{\theta}$ (but not its modulus).

From expression (\ref{cross-correl-def}), and using the formalism of Dolag and Bartelmann (1997), one can derive assuming weak lensing that 
\begin{eqnarray}
  \omega_{qg}(\theta) & = &
    {\displaystyle \frac{(s-1)}{\pi} \frac{3}{2}
    \left( \frac{H_o}{c} \right)^2 \Omega_m }
    {\displaystyle \int _{0}^{y _{\infty} }} dy
    {\displaystyle \frac { W_g(y) \, G_q(y) } {a(y)}}  \nonumber \\
    & &  \times
    {\displaystyle \int _{0}^{\infty }} dk \, k \,
    P_{gm}(k,y )\, J_0[f_K(y )k\theta ]  \, ,
         \label{analytic-cross-correl}
\end{eqnarray}
where $y$ is the comoving distance, which here parameterizes time
($y_{\infty}$ represents a redshift $z=\infty$),
and $k$ is the wavenumber of the density contrast in a plane wave
expansion;
$J_0$ is the zeroth-order Bessel function of first kind;
and $f_K(y)$ is the curvature-dependent radial distance ($=y$ for a flat
universe).
$P_{gm}(k,y )$ can be seen as the galaxy-mass cross-power spectrum (Jain et al. 2003), and under some assumptions (Guimar\~aes et al. 2001) may be expressed as 
$P_{gm}(k,y )=\sqrt{P_g(k) P_m(k,y )}$, where
$P_g(k)$ is the power spectrum for galaxies or galaxy groups and
$P_m(k,y)$ is the non-linear time evolved mass power spectrum.

M\'{e}nard et al. (2003) calculated the second order contribution to $\omega_{qg}(\theta)$ in addition to the dominant first order term given by expression (\ref{analytic-cross-correl}) and found that it can increase the amplitude of $\omega_{qg}(\theta)$ by 15\% to 20\% on scales below one degree.

\begin{figure}
  \centering
  \includegraphics[width=8.3 cm]{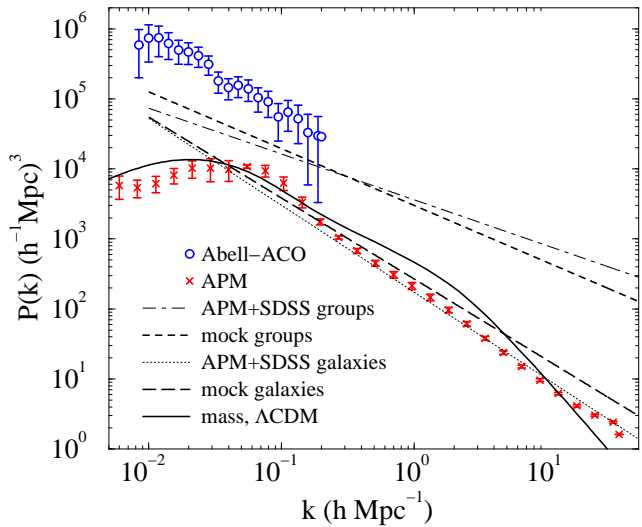}
  \caption{Power spectra for galaxies, groups, and mass distribution.
The mass power spectrum is for a concordance model.
The power law spectra for the APM+SDSS and mock catalogue sets are obtained from the respective auto-correlation function using the method described in the text.}
  \label{power-spectra}
\end{figure}

The mass power spectrum $P_m(k,y)$ for the $\Lambda$CDM model can be obtained analytically from the linear theory and the use of the non-linear prescription given by Peacock \& Dodds (1996).
Figure \ref{power-spectra} shows the mass power spectrum at $z=0$ obtained by the method described.

To obtain the three-dimensional power spectrum in real space for galaxies and groups $P_g(k)$ we use the corresponding two-point angular auto-correlation function $\omega_{gg}(\theta)$ and the relation (Peacock 1991)
\begin{equation}
  \omega_{gg}(\theta) = \int_0^\infty dy \, y^4 \phi^2(y)
  \int_0^\infty dk \frac k{2\pi} P_g(k) J_0(ky\theta) \;,
  \label{correl-power}
\end{equation}
where $\phi(y)$ is the selection function, normalized such that
$\int y^2 \phi dy=1$.
If we assume $\phi \propto y^{1/2} \exp [-(y/y_*)^2]$,
$\omega_{gg}(\theta) = B \theta^\beta$, and that the power spectrum is a power law, then expression (\ref{correl-power}) can be inverted, allowing us to find that
\begin{equation}
P_g(k) = \frac{\pi^3 2^{1+\frac{3}{2}\beta} y_*^{1-\beta} B}
{(\beta+2) \Gamma \left( -\frac{\beta}{2} \right)
\Gamma^2 \left( \frac{3}4 \right)}
k^{-(\beta+2)} \; .
\label{power-approx}
\end{equation}

We find that $y_*=400 h^{-1}$Mpc allows a good approximation for the selection function used in the generation of the galaxy mock catalogue.

We plot on Figure \ref{power-spectra} the power law spectra calculated from the auto-correlation functions shown in Figure  \ref{groupcorrel} using equation (\ref{power-approx}), and also the original spectra for APM galaxies (Gazta\~naga \& Baugh 1998) and Abell-ACO clusters (Miller \& Batuski 2001) for comparison.

\begin{figure}
  \centering
  \includegraphics[width=8.3 cm]{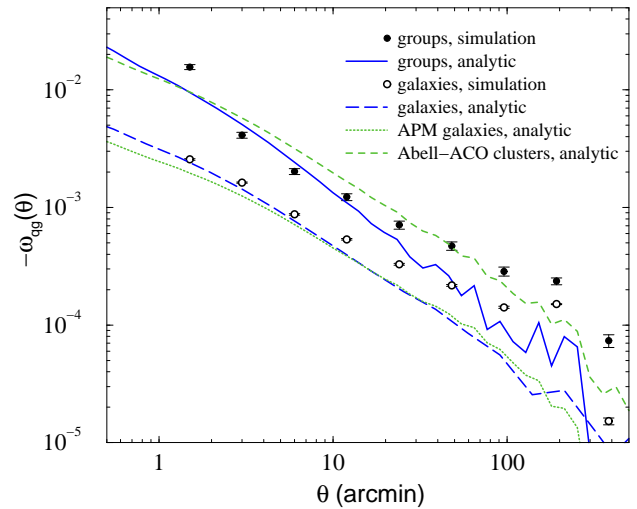}
  \caption{QSO-galaxy and QSO-group cross-correlation from simulation and analytic calculation. For the analytic results the only difference in the calculation is the lens (galaxy or group) power spectrum used.}
  \label{cross-correl_simulXanal}
\end{figure}

Figure \ref{cross-correl_simulXanal} shows the QSO-galaxy and QSO-group cross-correlation obtained from using Equation (\ref{analytic-cross-correl}) and the approximated power spectra for galaxies and groups obtained from the auto-correlation functions, using the method described in this section.
There is a general agreement between our mass-light simulation method results and the simple analytic cross-correlation results, which is a useful confirmation of the consistency of both approaches.

The analytic cross-correlation expression (\ref{analytic-cross-correl}) can be further simplified if we approximate the source and lens distributions by Dirac Delta Functions peaked at the average comoving distances 
(${\bar y}_q$ for the QSOs and ${\bar y}_g$ for the galaxies), and the power spectrum term by $P_{gm}(k)=Ak^\alpha$, then 
\begin{eqnarray}
  \omega_{qg}(\theta) & = &
  \frac{3\cdot2^\alpha (s-1)}{\pi} \left( \frac{H_o}{c} \right)^2
  \frac{\Gamma \left(\frac{\alpha}{2}+1\right)}{\Gamma \left(-\frac{\alpha}{2}\right)}
  \nonumber \\ 
  & &  \times    
  \frac{ f_K({\bar y}_q) -f_K({\bar y}_g)}{a({\bar y}_g)f_K({\bar y}_q)f_K^{\alpha +1}({\bar y}_g)} 
  \Omega_m A \theta^{-\alpha -2}  \, ,
  \label{acc-simply}
\end{eqnarray}
where $\theta$ is in radians.
If one assumes linear bias, then from expressions (\ref{acc-simply}) and (\ref{power-approx}) it can be deduced that the cross-correlation has the same angular dependence as the galaxy or group auto-correlation, that is to say, if $\omega_{gg}(\theta) \propto \theta^\beta$ then $\omega_{qg}(\theta)\propto \theta^\beta$.
This simple relation is roughly confirmed by the results (both from simulation and observation) shown in Figure \ref{groupcorrel} for the auto-correlation for galaxies and groups and in  Figure \ref{red-cross-correl} for the respective cross-correlations.

\section{Mass Estimation}
\label{mass_estim}

One simple way to estimate the mass of a lens is to assume a mass profile for it, calculate the expected magnification, and use expression (\ref{w_qg-mag-rel}) to determine the cross-correlation that a population of these lenses would generate (halo lens fitting).
One can then find the parameters for the mass profile chosen that best fit the observed QSO-lens cross-correlation, and with it the mass comprised by the lens.

This lens mass estimation method was used by Myers et al. (2003) in groups of galaxies identified in a two-dimensional sky projection, however these groups are not necessarily bound entities.
Also, the method essentially assumes that the lenses are identical and isolated objects, ignoring large-scale structure considerations such as clustering and filaments.
Nevertheless, the simplicity of the model is appealing, and its result can be attributed to some ``effective'' halo.
Our simulation allows us to test the validity of the method by comparing our mass estimates for lenses obtained from the QSO-lens cross-correlation and directly from the mock matter distribution.

We examine two popular choices for halo profiles, the singular isothermal sphere (SIS)  and the NFW profile (Navarro, Frenk and White 1997).

For a SIS halo profile $\rho_{SIS}(r)= \sigma_v /2\pi G r^2$, and
the mass inside a radius $r$ is $M_{SIS} = 2 \sigma_v^2 r / G$, where $\sigma_v$ is the velocity dispersion.

For the NFW halo profile
\begin{equation}
  \rho_{NFW}(r)= \frac{\delta_c \rho_c}{(r/r_s)(1+r/r_s)^2} \; ,
\end{equation}
where $\rho_c$ is the critical density.
We reduce the NFW profile to a one parameter description, the mass inside a 1.5$h^{-1}$Mpc radius sphere ($M_{1.5}$), using the relation (Maoz et al. 1997)
\begin{equation}
r_s= 0.3 \left( \frac{M_{1.5}}{10^{15}\msun} \right)^\lambda h^{-1} {\rm Mpc} \;
\end{equation}
where we use $\lambda=1/3$.

The convergence $\kappa(\theta)$ generated by a halo can be calculated projecting its mass density profile into a plane and using equation (\ref{convergence1}).
The shear can be obtained in this case of circular symmetry by equation (\ref{tangential_shear}).
See Wright \& Brainerd (2000) for explicit analytical expressions for lensing by halos.
The magnification can then be calculated using equation (\ref{magnification}).
\begin{figure}
  \centering
  \includegraphics[width=8.3 cm]{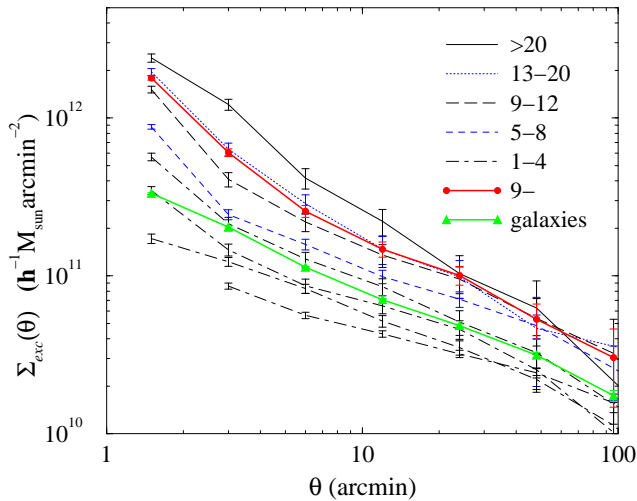}
  \vspace{0.cm}
  \caption{Excess surface density $\Sigma_{exc}$ around galaxies and groups in relation to the mean surface density of the universe to redshift 0.4 
($4.3\cdot10^{12}h^{-1}\msun{\rm arcmin}^{-1}$).
The legend numbers indicate the galaxy group membership. 
Errors are the standard deviation of the mean. 
}
  \label{mass_cone_group-rand}
\end{figure}

For the galaxies and galaxy groups in our mock catalogue we can estimate the lensing mass associated with them directly from the matter simulation.
Figure \ref{mass_cone_group-rand} shows the excess surface density $\Sigma_{exc}$ in a ring around galaxies and groups in relation to the mean surface density of the universe to redshift 0.4. 
The excess density is largely insensitive to redshift cuts above this value, since structures above it are not correlated to the luminous structures that we are interested.

The excess mass contained inside a redshift cone centred at a lens (obtained from an integration of the excess surface density) can be used to give an estimate of the average lens mass.
However this method in fact includes mass outside the nominal radius, since the cone angular aperture only excludes the mass outside the radius in the perpendicular direction of its axis.
The projected mass outside the radius in the line of sight direction is also included.
Therefore one would expect the mass obtained through this method to be an over-estimate of the real mass inside a chosen radius.
We show these masses around lens centres for cones with angular aperture corresponding to $1.5h^{-1}\rm Mpc$ radii for galaxies at an average redshift of 0.15 in Table \ref{mass-table}.

\begin{table*}
  \caption{Estimated average masses within $1.5h^{-1}\rm Mpc$ of the galaxy or group centre in units of $10^{14}h^{-1}\msun$. The columns are: 
$\sigma_v$ is the velocity dispersion for a SIS model and $\rm{M_{1.5, SIS}}$ is the corresponding estimated mass, 
$\rm{M_{1.5, NFW}}$ is the estimated mass in a NFW profile model, $\rm{M_{1.5, \Sigma exc}}$ is the estimated mass from an integrated excess surface mass density around galaxies or groups at the average redshift z=0.15 (errors are standard deviation from the mean), and $\rm{M_{1.5, sph}}$ is the average mass inside spheres centred at galaxy positions (errors are standard deviation).
$\rm{M_{1.5, \Sigma exc}}$ and $\rm{M_{1.5, sph}}$ are clearly only available for simulations.
Groups from simulations have 9 or more galaxies and groups from observation have 7 or more members.The observational data sets come from Myers et al. (2003 \& 2005). }
  \begin{center}
    \begin{tabular}{l  r@{}l  r@{.}l  r@{.}l  r@{.}l r@{.}l}
      \hline
      data set            & \multicolumn{2}{c}{$\sigma_v$ (km s$^{-1}$)} &  \multicolumn{2}{c}{$\rm{M_{1.5, SIS}}$}
      &  \multicolumn{2}{c}{$\rm{M_{1.5, NFW}}$}  &  \multicolumn{2}{c}{$\rm{M_{1.5, \Sigma exc}}$}  
      &  \multicolumn{2}{c}{$\rm{M_{1.5, sph}}$}   \\
      \hline
      galaxy, simulation  & 227&$^{+39}_{-48}$ & 0&4$\pm 0.1$   & 0&23$^{+0.20}_{-0.19}$ &   0&64$\pm 0.01$ &  0&11$\pm 0.10$ \\
      galaxy, observation & 470&$^{+59}_{-69}$ & 1&5$\pm 0.4$   & 1&3$\pm 0.5$           & \multicolumn{2}{c}{---} & \multicolumn{2}{c}{---}\\
      \hline
      group, simulation   & 409&$^{+46}_{-53}$ & 1&1$\pm 0.3 $  & 1&1$^{+0.6}_{-0.5}$    & 1&5$\pm 0.1$ & \multicolumn{2}{c}{---} \\
      group, observation  &1156&$^{+93}_{-327}$& 9&$^{+3}_{-4}$ & 12&$\pm 9$             & \multicolumn{2}{c}{---} & \multicolumn{2}{c}{---}\\
      \hline
    \end{tabular}
  \end{center}
  \label{mass-table}
\end{table*}

Table \ref{mass-table} compares the masses associated with galaxies and groups using halo lens fitting to the simulation and observational QSO-group cross-correlation results and directly from the matter distribution simulation.

For galaxies the redshift is well defined (to projected galaxy groups it is not) and therefore we can estimate the mass inside a sphere centred on the galaxy coordinates.
We calculate this mass for a $1.5h^{-1}$Mpc radius from the centre of the lenses. 
Such a large radius for a galaxy is chosen to make possible the comparison with similar measures for groups and therefore should be seen as a measure not of the individual galaxies but of the galactic environment.

Looking  across the row in Table \ref{mass-table} for the simulation results having galaxies as lens centres, we note that the average mass inside a sphere of radius $1.5h^{-1}$Mpc  is roughly six times smaller than the mass computed from the excess surface density. 
The masses estimated from the simple halo profile lensing models fall between these two.
Assuming that the same relation between the described masses applies for groups, then the estimated masses from the halo lens fittings are in reasonable agreement (tending to be an overestimate) with the average real mass associated with groups.

The vertical comparison of the values in the three first columns in Table \ref{mass-table} also reflects the significant amplitude disagreement discussed in Section \ref{Mag_centre} between observation and simulation, but to a less dramatic degree than the power law analysis.
For the galaxy results the mass inferred from observations using the halo profile lensing method is $\sim$5 times larger than the corresponding results from the simulation, and for groups this same factor is $\sim$10.

\section{Discussion}
\label{conclusion}

We have pursued a computational approach to the problem of QSO-group cross-correlation, which uses the same galaxy group identification procedure of Myers et al. (2003) in galaxy survey data, who found a large anti-correlation signal.
The method consists in simulating the mass and light distribution, and directly computing the magnification due to gravitational lensing caused by the mass concentrations traced by selected galaxies. 
The background-foreground correlation is computed directly from the average angular magnification around the group centres, while previous works that also use simulations calculate the cross-correlation from lensing field statistics.
Our approach takes into account the large scale structure of the universe, deviations from the weak lensing approximation, the actual galaxy distribution, and selection criteria.

The large scale structure of the universe implies that the groups are clustered, and therefore this feature is properly taken into account in our calculations, in contrast to a simpler approach that models lenses as isolated objects, as used by Myers et al. (2003).
Nevertheless, this simple approach was shown to be roughly adequate for lens mass estimation.
The mass obtained from isolated halo fitting to cross-correlation results obtained using mock galaxies drawn from a suitably biased Hubble Volume simulation  are comparable for the two halo profiles examined, and the direct mass estimate from the mock matter distribution.

The consideration of the LSS also allows the inclusion of structures that are not connected to the groups, but act as noise generators for lensing measurements.
In fact, most of the lensing for distant sources come from structures around half of the way to the observer, and these structures are not connected to the galaxy groups since they are much closer to the observer.
This background lensing signal from uncorrelated structures is the main noise source for the lensing signal from structures associated with observable galaxies.

Our results for galaxy groups predict a strong dependence of the cross-correlation on the group galaxy membership, which is in accord with the calculated mass estimates for these groups.
The cross-correlation curves suggest effective mass profiles for the groups that are distinct in amplitude, but comparable in slope, just favouring slightly cuspier profiles for more massive structures.

The results for galaxy groups of membership 1, i.e. isolated (field) galaxies, when compared to the results of the average galaxy suggest an interesting application. Both the QSO-galaxy cross-correlations and the direct mass estimates for the two populations of lenses indicates that the average galaxy is twice as massive as field galaxies.
If the galaxy population is large enough, then it could in practice be broken down into smaller sub-sets according to a chosen characteristic and the cross-correlation observation could be used to compare the relative masses of these sub-populations.

Our simulation lacks mass resolution.
A higher resolution simulation of the matter distribution at small redshifts would allow a more precise determination of deviations from linearity of the weak lensing approximation.
It would also allow the investigation of smaller angular scales, including halo substructure and the magnification closer to the core of galaxies and clusters.
It is possible that substructure may make non-linear magnification important even at regions far from the identified galaxy or galaxy group centre.
Therefore a higher mass resolution simulation could yield a stronger expected cross-correlation due to non-linear lensing, not only for small $\theta$ but also for larger angles.
This possible effect is worth further investigation.

The persisting discrepancy between the amplitude of the expected QSO-galaxy and QSO-group cross-correlations in relation to the much lower values expected from the cosmological and lensing models adopted is a stimulus for further investigations.

It may be possible that there are observational issues that cause a systematic error in the cross-correlation determination.
The hypothesis of dust is by now very weakened, since the reddening of lensed QSO was found not to be significant by Myers et al. (2003), and the strong positive cross-correlation measured by some groups when using bright QSOs also argues against it.
The dust would need to obscure only faint QSOs and cause no reddening to be a viable explanation.

On the other hand, if all the observational issues are very well understood, and the cross-correlation signal due to lensing is stronger than predictions such as we provide in this paper, then various possibilities are indicated.
One possibility is a higher mass density for the universe, which would increase the overall weighting of the lenses (the lensing factor). 
This possibility is highly constrained from other, probably more precise and accurate determinations of $\Omega_m$.
Another possibility is that the lenses themselves (galaxies and clusters) are more massive than the $\Lambda$CDM model predicts.
A third option is that the lensing efficiency is higher than expected from our simulation or analytical calculations that assume weak lensing due to a more prominent role of non-linear lensing. At high density regions (near the core or due to substructure) the magnification is very non-linear and could in principle give a higher contribution to the signal than what is currently being simulated.
This last possibility is favored by our observation that the disagreement is larger for galaxy groups than for individual galaxies. Since groups trace regions of higher density than isolated galaxies, it is expected that non-linear effects would be more pronounced for groups.

\section*{Acknowledgments}
We thank Shaun Cole for assistance with the Hubble Volume Simulation output and galaxy mock generation, and Phil Outram for useful discussions and comments.

\onecolumn
\appendix

\section{Cross-Correlation Definition and Estimator}
\label{cross-corr-def-est}

Here we show the equivalence between expressions (\ref{estimator}) and (\ref{cross-correl-def}) for the angular cross-correlation between elements belonging to two populations.

We can describe the populations of objects by two sets of position vectors
${\bm\alpha}_i$ ($i=1,N_\alpha$) and  ${\bm\beta}_j$ ($j=1,N_\beta$), where
$N_{\alpha,\beta}$ is the number of elements in either set.
So the population density $n_\alpha$ can be written as
\begin{equation}
n_\alpha (\phi) = \sum_{i=1}^{N_\alpha}{\delta_D(\phi-\alpha_i)} \;,
\end{equation}
where $\delta_D$ is a Dirac delta function, and the population mean suface density $\bar{n}_\alpha$ is
\begin{equation}
\bar{n}_\alpha \equiv \left< n_\alpha(\phi) \right>
\equiv  \frac{\int {n_\alpha(\phi) d\phi}} {\int{d\phi}}
= \frac {N_\alpha}{A}  \;,
\end{equation}
where $A$ is the survey area, and a similar expression holds for $n_\beta$.

Therefore the cross-correlation between populations $\alpha$ and $\beta$ is, according to expression (\ref{cross-correl-def}),
\begin{eqnarray}
  \omega_{\alpha \beta}(\theta) &=&
  \left<
  \left( \sum_{i=1}^{N_\alpha} \frac{\delta_D({\bm\phi}-{\bm\alpha}_i)}
       {\bar{n}_\alpha} -1 \right)
  \left( \sum_{j=1}^{N_\beta} \frac{\delta_D({\bm\phi}+{\bm\theta}-{\bm\beta}_j)}
       {\bar{n}_\beta} -1 \right)
  \right> \\
 &=& \frac 1{\bar{n}_\alpha \bar{n}_\beta}
  \left< \sum_{i}\sum_{j} \delta_D({\bm\phi}-{\bm\alpha}_i) \delta_D({\bm\phi}+{\bm\theta}-{\bm\beta}_j) \right>
  - \frac 1{\bar{n}_\alpha} \left< \sum_{i} \delta_D({\bm\phi}-{\bm\alpha}_i)\right>
  - \frac 1{\bar{n}_\beta} \left< \sum_{j} \delta_D({\bm\phi}+{\bm\theta}-{\bm\beta}_j)\right>
  + 1  \nonumber \\
  &=& \frac 1{\bar{n}_\alpha \bar{n}_\beta}
  \frac{\sum_{ij} \delta_D\left(\theta-|{\bm\beta}_j-{\bm\alpha}_i|\right)}
  {2\pi\theta A} - 1 -1 + 1  \nonumber \\
  &=& \frac {DD(\theta)}{DR(\theta)} - 1  \nonumber \;,
\end{eqnarray}
which is expression (\ref{estimator}).

The last step takes two facts into consideration; that $DD(\theta)$ is the number of pairs of elements from the two populations such that their separation
$|{\bm\beta}_j-{\bm\alpha}_i|$ is $\theta$, and that for a random distribution of both populations $DR(\theta)=2\pi\theta A \bar{n}_\alpha \bar{n}_\beta $.

\end{document}